\documentclass[preprint,aps,prd,showpacs,nofootinbib]{revtex4}
\usepackage{mathrsfs}
\usepackage{amsmath}
\usepackage{graphicx}
\usepackage{subfigure}

\newcommand{\bea}{\begin{eqnarray}}
\newcommand{\eea}{\end{eqnarray}}
\newcommand{\beq}{\begin{equation}}
\newcommand{\eeq}{\end{equation}}

\def\/{\over}
\makeatletter

\newcommand{\Rmnum}[1]{\expandafter\@slowromancap\romannumeral #1@}
\makeatother

\begin{document}

\title{The Lamb shift in de Sitter spacetime}
\author{  Wenting Zhou$^{1}$ and Hongwei Yu$^{1,2,}$\footnote{Corresponding author}  }
\affiliation{$^1$ Department of Physics and Key Laboratory of Low
Dimensional Quantum Structures and Quantum
Control of Ministry of Education,\\
Hunan Normal University, Changsha, Hunan 410081, China \\
$^2$ Center for Nonlinear Science and Department of Physics, Ningbo
University,  Ningbo, Zhejiang 315211, China}

\begin{abstract}
We study the Lamb shift of both  freely-falling and  static
two-level atoms in interaction with  quantized conformally coupled
massless scalar fields in the de Sitter-invariant vacuum. We find
that  the Lamb shifts of both  freely-falling and  static atoms are
in structural similarity to that of an inertial atom immersed in a
thermal bath in a Minkowski spacetime. For the freely-falling atom,
the Lamb shift gets a correction  as if it was immersed in a thermal
bath at the Gibbons-Hawking temperature, thus
 revealing clearly the intrinsic thermal nature of de Sitter spacetime.
For the static atom, the Lamb shift is affected by a combination of
the effect of the intrinsic thermal nature of de Sitter spacetime
and the Unruh effect associated with the inherent acceleration of
the atom.

\end{abstract}
 \maketitle

\baselineskip=16pt

\section{Introduction}
The Lamb shift is one of the most remarkable observable phenomena in
physics which has been precisely measured in experiment and it has
attracted a great deal of interest since its discovery in
1947~\cite{Lamb47}. So far, the Lamb shift has been investigated in
various circumstances, in the presence of
cavities~\cite{Meschede90}, or in a thermal
bath~\cite{Barton72,Knight72,Farley and Wing81}, for example. It has
recently been shown that the non-inertial motion of the atom also
induces corrections to the Lamb
shift~\cite{Audretsch95,Passante98,L.Rizzuto07,ZhuYu10}. However,
all the aforementioned studies are concerned with flat spacetimes.
Therefore, it remains interesting to see what happens if the atom is
placed in a curved spacetime rather than a flat one. In this regard,
it is interesting to note that the Lamb shift has recently been
calculated of static atoms in interaction with fluctuating vacuum
massless scalar fields in a curved background, i.e., the exterior of
Schwarzschild black hole, and it is found that the Lamb shift gets
corrected as a result of both scattering of vacuum field modes off
the spacetime curvature and the Hawking radiation from the black
hole~\cite{ZhouYu10}.

In the present paper, we plan to calculate the Lamb shift in a de
Sitter spacetime which describes an empty universe with a positive
cosmological constant. It is the unique maximally symmetric curved
spacetime and enjoys an important status in curved spacetimes just
as that of the Minkowski spacetime in flat spacetimes.   Our
interest in this issue are twofold. First, our universe is believed,
according to the current observations and the inflation theory, to
approach  de Sitter spacetime in the far past and the far future.
Second, there may exist a holographic duality between quantum
gravity on de Sitter spacetime and a conformal field theory living
on the boundary identified with the timelike infinity of de Sitter
spacetime~\cite{DSCFT}. So, it is interesting  to investigate the
Lamb shift in this special curved spacetime  and this is exactly
what we plan to do in the present paper. Using an elegant formalism
suggested by Dalibard, Dupont-Roc and
Cohen-Tannoudji(DDC)~\cite{Dalibard82,Dalibard84}, which allows a
separation of the contributions of vacuum fluctuations and the
radiation reaction to the energy shifts,  we will calculate  the
Lamb shift of both a freely falling atom and a static one with an
inherent acceleration in interaction with vacuum fluctuations of
quantized massless conformally coupled scalar fields in de Sitter
spacetime. Let us note that the quantization of  scalar fields in
this spacetime has been extensively studied in the literature
\cite{QFT,Bunch and
Davies,Tagirov,Ford,Schomblond,Mottola,Allen,Allen87,Polarski,Polarski
prd}.

When referring to the vacuum fluctuations in quantum field theory,
we should first specify the vacuum states. Our research is done in
the de Sitter-invariant vacuum state which is  deemed to a natural
vacuum in this spacetime since  it preserves the de Sitter
invariance~\cite{Allen} and therefore enjoys a special status as
that of the Minkowski vacuum in  flat spacetimes. Our calculations
show that the Lamb shift for an atom moving on a timelike geodesic
(freely falling) is identical to that of an inertial one immersed in
a thermal bath at the Gibbons-Hawking temperature. For a static
atom, the Lamb shift is modified as opposed to that in a flat
spacetime by the combined effects of the intrinsic thermal nature of
de Sitter spacetime characterized by the Gibbons-Hawking temperature
and the atomic inherent acceleration.

\section{the general formalism}
We consider a pointlike two-level atom interacting with quantized
conformally coupled massless scalar field in de Sitter spacetime.
The two stationary atomic eigenstates are represented by $|+\rangle$
and $|-\rangle$ and their corresponding energies are
$\frac{1}{2}\omega_0$ and $-\frac{1}{2}\omega_0$ respectively. The
Hamiltonian that determines the evolution of the atom-field system
with respect to the proper time of the atom, $\tau$, is given by,
 \beq
H=H_A(\tau)+H_F(\tau)+H_I(\tau)\;.
 \eeq
Here $H_A(\tau)$ is the Hamiltonian of the atom. In Dicke's notation
\cite{Dicke}, it is given by
 \beq
H_A(\tau)=\omega_0R_3(\tau)\label{atomic Hamiltonian}
 \eeq
with
$R_3(0)=\frac{1}{2}|+\rangle\langle+|-\frac{1}{2}|-\rangle\langle-|$
; $H_F(\tau)$ is the free Hamiltonian of the quantum scalar field
 \beq
H_F(\tau)=\int
   d^3k\;\omega_{\vec{k}}\;a^+_{\vec{k}}\;
   a_{\vec{k}}\frac{dt}{d\tau}\;,
 \eeq
in which $a^+_{\vec{k}}$ and $a_{\vec{k}}$ are the creation and
annihilation operators with momentum $\vec{k}$\;; $H_I(\tau)$ is the
Hamiltonian that describes the interaction between the atom and the
field,
 \beq
H_I(\tau)=\mu R_2(\tau)\phi(x(\tau))\;,
 \eeq
where $\mu$ is a small coupling constant that is assumed to be
small, $R_2(0)=\frac{1}{2}i[R_-(0)-R_+(0)]$, and
$R_+(0)=|+\rangle\langle-|$, and $R_-(0)=|-\rangle\langle+|$ are the
atomic raising and lowering operators. These operators obey the
angular momentum algebra: $[R_3,R_\pm]=\pm R_\pm$, $[R_+,R_-]=2R_3$.
$\phi(x)$ is the scalar field operator in de Sitter spacetime and it
satisfies the wave equation
 \beq
(\nabla_\mu\nabla^\mu+m^2+\xi R)\;\phi(x)=0\;,
 \eeq
where $m$ is the mass of the scalar field, $\xi$ is the coupling
constant, and $R$ is the Ricci scalar curvature of the spacetime. In
the case of conformally coupled scalar field in a four dimensional
spacetime here, $\xi=1/6$. The coupling is effective only on the
atomic trajectory $x(\tau)$.

From the above Hamiltonians, we can derive the Heisenberg equations
of the atomic and the field's variables and their solutions can then
be divided into two parts: a free part which exists even when there
is no coupling between the atom and the field; a source part that is
caused by the coupling between the two and is characterized by the
small coupling constant $\mu$. Assuming the original state of the
field is the de Sitter-invariant vacuum state, i.e., the Euclidean
or Bunch-Davies vacuum state \cite{Bunch and Davies}, choosing a
symmetric operator ordering between the atomic and the field's
variables, and proceeding in a manner similar to that in
Ref.~\cite{Dalibard84,Audretsch95}, we can identify the following
effective Hamiltonians  to the order $\mu^2$,
 \bea
H^{eff}_{vf}(\tau)&=&\frac{1}{2}i\mu^2\int^{\tau}_{\tau_0}d\tau'C^F(x(\tau),x(\tau'))
[R^f_2(\tau'),R_2^f(\tau)]\;,\label{Heffvf}\\
H^{eff}_{rr}(\tau)&=&-\frac{1}{2}i\mu^2\int^{\tau}_{\tau_0}d\tau'\chi^F(x(\tau),x(\tau'))
\{R^f_2(\tau'),R_2^f(\tau)\}\;,\label{Heffrr}
 \eea
the sum of which governs the time evolution of the atomic
observables. Here $\{\;,\;\}$ and $[\;,\;]$ denote the commutator
and anticommutator respectively. $C^F(x(\tau),x(\tau'))$ and
$\chi^F(x(\tau),x(\tau'))$ are separately the symmetric correlation
function and the linear susceptibility of the field, and they are
defined as
 \bea
C^F(x(\tau),x(\tau'))&=&\frac{1}{2}\langle0|\{\phi^f(x(\tau)),\phi^f(x(\tau'))\}
|0\rangle\;,\\
\chi^F(x(\tau),x(\tau'))&=&\frac{1}{2}\langle0|[\phi^f(x(\tau)),\phi^f(x(\tau'))]
|0\rangle\;.
 \eea
Taking the expectation values of Eqs.~(\ref{Heffvf}) and
(\ref{Heffrr}) on a generic atomic state $|b\rangle$ yields the
contributions of vacuum fluctuations and the radiation reaction to
the energy shift of level $b$
 \bea
(\delta E_b)_{vf}&=&-i\mu^2\int_{\tau_0}^\tau
   d
   \tau'C^F(x(\tau),x(\tau'))\;\chi^A_{b}(\tau,\tau')\;,\label{general delta vf}\\
(\delta E_b)_{rr}&=&-i\mu^2\int_{\tau_0}^\tau
   d
   \tau'\chi^F(x(\tau),x(\tau'))\;C^A_{b}(\tau,\tau')\;,\label{general delta rr}
 \eea
in which $C^A_b(\tau,\tau')$ and $\chi^A_b(\tau,\tau')$ are,
separately, the symmetric correlation function and the linear
susceptibility of the atom. They are defined as
 \bea
C^A_{b}(\tau,\tau')&=&\frac{1}{2}\langle
b|\{R_2^f(\tau),R_2^f(\tau')\}|b\rangle\;,\\
\chi^A_{b}(\tau,\tau')&=&\frac{1}{2}\langle
b|[R_2^f(\tau),R_2^f(\tau')]|b\rangle\;.
 \eea
They do not depend on the atomic trajectory and are determined by
the internal structure of the atom itself. Their explicit forms are
 \bea
C^A_{b}(\tau,\tau')&=&\frac{1}{2}\sum_{d}|\langle
b|R_2(0)|d\rangle|^2\,
   (e^{i\omega_{bd}\Delta\tau}+e^{-i\omega_{bd}\Delta\tau})\;,\\
\chi^A_{b}(\tau,\tau')&=&\frac{1}{2}\sum_{d}|\langle
b|R_2(0)|d\rangle|^2\,
   (e^{i\omega_{bd}\Delta\tau}-e^{-i\omega_{bd}\Delta\tau})\;.
 \eea
Here $\omega_{bd}=\omega_{b}-\omega_{d}$ and the sum extends over a
complete set of the atomic eigenstates.

\section{The Lamb shift of a freely falling atom in de Sitter spacetime}

In this section, we consider the Lamb shift of a freely-falling atom
in interaction with a quantized conformally coupled massless scalar
field in de Sitter spacetime. There are several different coordinate
systems that can be chosen to parameterize  de Sitter
spacetime~\cite{QFT}. Here we choose to work with the global
coordinate system $(t,\chi,\theta,\phi)$ under which the
freely-falling atom is comoving with the expansion. The line element
is
 \beq
ds^2=dt^2-\alpha^2\cosh^2(t/\alpha)[d\chi^2+\sin\chi^2(d\theta^2+\sin^2\theta
     d\varphi^2)]\;
 \eeq
with $\alpha=3^{1/2}\Lambda^{-1/2}$, where $\Lambda$ is the
cosmological constant. The parameter $t$ is often called the world
or cosmic time. The scalar curvature of the spacetime is
$R=12\alpha^{-2}$. The canonical quantization of a massive scalar
field with this metric has been done in
Ref.~\cite{Allen,Allen87,Schomblond,Bunch and Davies}. In the global
coordinates, solve the equation of motion to get a complete set of
eigenmodes and define a de Sitter-invariant vacuum, then the
Wightman function of the massive scalar field can be
found~\cite{Allen87}
 \beq
G^+(x(\tau),x(\tau'))=-\frac{1}{16\pi\alpha^2}\frac{\frac{1}{4}-\nu^2}
{cos(\pi\nu)}F\biggl(\frac{3}{2}+\nu,\frac{3}{2}-\nu\;;2\;;\frac{1-Z(x,x')}{2}\biggr)\;
 \eeq
with $F$ being the hypergeometric function and
 \bea
Z(x,x')&=&\sinh{\frac{t}{\alpha}}\sinh{\frac{t'}{\alpha}}-
\cosh{\frac{t}{\alpha}}\cosh{\frac{t'}{\alpha}}\cos{\Omega}\;,\nonumber\\
\cos{\Omega}&=&\cos{\chi}\cos{\chi'}+\sin{\chi}\sin{\chi'}
[\cos{\theta}\cos{\theta'}+\sin{\theta}\sin{\theta'}\cos{(\varphi-\varphi')}]\;,\nonumber\\
\nu&=&\sqrt{\frac{9}{4}-\frac{12}{R}(m^2+\xi R)}\;.
 \eea
In the massless and conformal coupling limit, the Wightman function
for a freely-falling atom can be simplified to be
 \beq
G^+(x(\tau),x(\tau'))=-\frac{1}{16\pi^2\alpha^2\sinh^2
   (\frac{\tau-\tau'}{2\alpha}-i\epsilon)}\;.
 \eeq
This leads to the following two statistical functions of the field
 \bea
C^F(x(\tau),x(\tau'))&=&-\frac{1}{32\pi^2\alpha^2}
     \biggl[\frac{1}{\sinh^2(\frac{\tau-\tau'}{2\alpha}-i\epsilon)}+
     \frac{1}{\sinh^2(\frac{\tau-\tau'}{2\alpha}+i\epsilon)}\biggr]\;,\\
\chi^F(x(\tau),x(\tau'))&=&\frac{i}{4\pi\cos(\frac{i(\tau-\tau')}
     {2\alpha})}\delta'(\tau-\tau')\;.
 \eea
Inserting them into Eqs.~(\ref{general delta vf}) and (\ref{general
delta rr}) and assuming the proper time interval
$\Delta\tau=\tau-\tau'$ to be sufficiently long, we can calculate
the contributions of vacuum fluctuations and the radiation reaction
to the energy shift of level $b$. Here, it is not easy to calculate
the involved integrals  using the  residual theorem and contour
integration techniques. So, we first perform the Fourier transform
on the above two statistical functions  to obtain
 \bea
C^F(x(\tau),x(\tau'))&=&\frac{1}{8\pi^2}\int_{0}^{\infty}
   d\omega\;\omega\coth(\pi\alpha\omega)
   (e^{i\omega\Delta\tau}+e^{-i\omega\Delta\tau})\;,\\
\chi^F(x(\tau),x(\tau'))&=&-\frac{1}{8\pi^2}\int_{0}^{\infty}d\omega\;\omega
   (e^{i\omega\Delta\tau}-e^{-i\omega\Delta\tau})\;.
 \eea
Then the contributions of vacuum fluctuations and the radiation
reaction can be expressed respectively as
 \bea
(\delta E_b)_{vf}&=&-\frac{i\mu^2}{16\pi^2}\sum_{d}|\langle
b|R_2(0)|d\rangle|^2\times\nonumber\\&&\quad\;
          \int_{0}^{\infty}d\omega\int_{0}^{\infty}d\Delta\tau\;
          \omega\coth(\pi\alpha\omega)\;
          (e^{i\omega\Delta\tau}+e^{-i\omega\Delta\tau})
          (e^{i\omega_{bd}\Delta\tau}-e^{-i\omega_{bd}\Delta\tau})\;,
 \eea
 and
 \bea
(\delta E_b)_{rr}&=&\frac{i\mu^2}{16\pi^2}\sum_{d}|\langle
b|R_2(0)|d\rangle|^2\times\nonumber\\&&\quad\;\quad\;
          \int_{0}^{\infty}d\omega\int_{0}^{\infty}d\Delta\tau\;\omega
          (e^{i\omega\Delta\tau}-e^{-i\omega\Delta\tau})
          (e^{i\omega_{bd}\Delta\tau}+e^{-i\omega_{bd}\Delta\tau})\;.
          \quad\;\quad\;\quad\;
 \eea
Further simplification gives
 \bea
 (\delta
E_b)_{vf}&=&\frac{\mu^2}{8\pi^2}\sum_{d}|\langle
          b|R_2(0)|d\rangle|^2\int_{0}^{\infty}d\omega\;P
          \biggl(\frac{\omega}{\omega+\omega_{bd}}-
          \frac{\omega}{\omega-\omega_{bd}}\biggr)\coth(\pi\alpha\omega)\;,\\
(\delta E_b)_{rr} &=&-\frac{\mu^2}{8\pi^2}\sum_{d}|\langle
b|R_2(0)|d\rangle|^2
          \int_{0}^{\infty}d\omega\;
          P\biggl(\frac{\omega}{\omega+\omega_{bd}}+\frac{\omega}{\omega-\omega_{bd}}\biggr)\;.
 \eea
Here and after $P$ denotes the Principal Value. So, for the
contribution of vacuum fluctuations to the energy shifts of the two
levels, we have
 \bea
&&(\delta
E_+)_{vf}=\frac{\mu^2}{32\pi^2}\int_{0}^{\infty}d\omega\;
          P\biggl(\frac{\omega}{\omega+\omega_{0}}
          -\frac{\omega}{\omega-\omega_{0}}\biggr)\coth(\pi\alpha\omega)\;,
          \label{deltaE+vf}\\&&
(\delta
E_-)_{vf}=\frac{\mu^2}{32\pi^2}\int_{0}^{\infty}d\omega\;
          P\biggl(\frac{\omega}{\omega-\omega_{0}}
          -\frac{\omega}{\omega+\omega_{0}}\biggr)\coth(\pi\alpha\omega)\;.
          \label{deltaE-vf}
 \eea
In the above computation, we have used the relation $\sum_d|\langle
b|R_2(0)|d\rangle|^2=1/4$. The contribution of vacuum fluctuations
to the shifts of two different levels differs only in sign. For the
contribution of the radiation reaction, we find
 \beq
(\delta E_+)_{rr}=(\delta
E_-)_{rr}=-\frac{\mu^2}{32\pi^2}\int_{0}^{\infty}d\omega
P\biggl(\frac{\omega}{\omega+\omega_{0}}+\frac{\omega}{\omega-\omega_{0}}\biggr)\;.
 \eeq
Obviously, the radiation reaction contributes the same to each
energy level's shift and it is equal to that in a four dimensional
Minkowski spacetime.

Adding up the contributions of vacuum fluctuations and  the
radiation reaction, we obtain the total energy shifts,
 \bea
&&\delta E_+=\frac{\mu^2}{16\pi^2}\int_{0}^{\infty}d\omega\;
          \biggl[\frac{\omega}{\omega+\omega_{0}}
          \frac{1}{e^{2\pi\alpha\omega}-1}
          -\frac{\omega}{\omega-\omega_{0}}
          \biggl(1+\frac{1}{e^{2\pi\alpha\omega}-1}\biggr)\biggr]\;,\\&&
\delta E_-=\frac{\mu^2}{16\pi^2}\int_{0}^{\infty}d\omega\;
          \biggl[\frac{\omega}{\omega-\omega_{0}}
          \frac{1}{e^{2\pi\alpha\omega}-1}
          -\frac{\omega}{\omega+\omega_{0}}
          \biggl(1+\frac{1}{e^{2\pi\alpha\omega}-1}\biggr)\biggr]\;.
 \eea
 The relative energy shift, i.e., the Lamb shift, is given
by $\Delta=\delta E_+-\delta E_-$, or by $\Delta=(\delta
E_+)_{vf}-(\delta E_-)_{vf}$ directly as the radiation reaction
contributes the same to each level's shift. So, the Lamb shift in de
Sitter space is entirely caused by vacuum fluctuations and is found
to be
 \beq
\Delta=\Delta_0+\Delta_{T_f}
 \eeq
with
 \bea
\Delta_0&=&\frac{\mu^2}{16\pi^2}\int_{0}^{\infty}d\omega\;
         \biggl(\frac{\omega}{\omega+\omega_{0}}-\frac{\omega}
         {\omega-\omega_{0}}\biggr)\;,\label{shift in Minkowski spacetime}\\
\Delta_{T_f}&=&\frac{\mu^2}{8\pi^2}\int_{0}^{\infty}d\omega
         P\biggl(\frac{\omega}{\omega+\omega_{0}}-\frac{\omega}
         {\omega-\omega_{0}}\biggr)
         \frac{1}{e^{2\pi\alpha\omega}-1}\;.\label{Tfreely falling}
 \eea
Here $\Delta_0$ is just the Lamb shift of an inertial two-level atom
in a free Minkowski spacetime with no boundaries. It is
logarithmically divergent and this divergence  is expected for a
non-relativistic treatment as what we do here. However, we can
remove the divergence by introducing a cutoff  on the upper limit of
the integration.  A reasonable cutoff frequency, $\omega_{max}$, was
suggested by Bethe who  took it to be $m$ (in SI units, it is
$mc^2/\hbar$) with $m$ being the mass of electron
\cite{H.A.Bethe47,Welton48}. Noticeably, the divergence can also be
removed, if one resorts to a fully relativistic
approach~\cite{French-Weisskopf49,Kroll-Lamb49}. Note that in the
second term, $\Delta_{T_f}$, a thermal factor
$(e^{2\pi\alpha\omega}-1)^{-1}$ appears. This term is similar to the
correction to that of an inertial atom immersed in a thermal bath in
a Minkowski spacetime at the temperature
$T_f=1/2\pi\alpha$~\cite{Barton72,Knight72,Farley and Wing81}. So,
for the freely-falling atom in de Sitter spacetime, the Lamb shift
is revised, as opposed to that in a flat unbounded spacetime, by a
thermal-like term as if it was immersed in a thermal bath at the
temperature $T_f=1/2\pi\alpha$, which is exactly the Gibbons-Hawking
temperature. We recover, in terms of the Lamb shift, Gibbons and
Hawking's result that reveals the thermal nature of de Sitter
spacetime~\cite{Gibbons77}. Finally, let us note that since our
universe is presumably in a phase of accelerating expansion and may
approach a de sitter space in the future, one may wonder whether the
correction to the Lamb shift due to the spacetime being de Sitter
rather Minkowski is experimentally measurable. In this regard, it is
worth noting that the cosmological constant is, according to current
observations, very tiny, corresponding to a Gibbons-Hawking
temperature of only $\sim10^{-30}K$. So, the correction  is
insignificant and it is therefore unrealistic for any actual
experimental measurement.

\section{The Lamb shift of a static atom in de Sitter spacetime}

For a static atom, we choose to work in the static coordinate system
in which the line element is
 \beq
ds^2=\biggl(1-\frac{r^2}{\alpha^2}\biggr)d^2\tilde{t}
    -\biggl(1-\frac{r^2}{\alpha^2}\biggr)^{-1}dr^2-
     r^2(d\theta^2+\sin^2\theta d\varphi^2)\;.
 \eeq
The metric possesses a coordinate singularity $r=\alpha$. The origin
$r=0$ corresponds to the position of the observer, so the
singularity is the event horizon for him. Just like the Rindler
wedge in a flat spacetime, the coordinates
$(\tilde{t},r,\theta,\varphi)$ only cover part of de Sitter
spacetime. The worldlines of constant $r$ are uniformly accelerated
timelike curves and only $r=0$ is a geodesic (with proper time $t$).
The static and the global coordinates are related by
 \beq
r=\alpha\cosh(t/\alpha)\sin\chi,\quad\;\tanh(\tilde{t}/\alpha)=\tanh(t/\alpha)\sec\chi\;.
 \eeq
Obviously, the worldline $r=0$ in the static coordinate coincides
with the worldline $\chi=0$ in the global coordinate and an atom at
rest with $r\neq0$ in the static coordinate will be accelerated
relative to the observer at rest in the global coordinate with
$\chi=0$.

Similarly, in the static coordinate system, one can find out a set
of complete eigenmodes by solving the field
equation~\cite{Polarski,Polarski prd} and define a de
Sitter-invariant vacuum. Then the Wightman function for the massless
conformally coupled scalar field is~\cite{Polarski717,Gal'tsov}
 \beq
G^+(x(\tau),x(\tau'))=-\frac{1}{8\pi^2\alpha^2}
    \frac{\cosh(\frac{r^*}{\alpha})\cosh(\frac{{r^*}'}{\alpha})}
    {\cosh(\frac{\tilde{t}-\tilde{t}'}{\alpha}-i\epsilon)-\cosh(\frac{r^*-{r^*}'}{\alpha})}\;
 \eeq
with $r^*=\frac{\alpha}{2}\ln\frac{\alpha+r}{\alpha-r}$. For a
static atom, it can be simplified to be
 \beq
G^+(x(\tau),x(\tau'))=-\frac{1}{16\pi^2\kappa^2\sinh^2(\frac{\tau-\tau'}{2\kappa}-i\epsilon)}
 \eeq
with $\kappa=\sqrt{g_{00}}\;\alpha$\;. In deriving  the above
result, we have used the relation,
$\Delta\tau=\sqrt{g_{00}}\;\Delta\tilde{t}$.
  Two statistical functions of the field are then easily obtained
 \bea
C^F(x(\tau),x(\tau'))&=&-\frac{1}{32\pi^2\kappa^2}
     \biggl[\frac{1}{\sinh^2(\frac{\tau-\tau'}{2\kappa}-i\epsilon)}+
     \frac{1}{\sinh^2(\frac{\tau-\tau'}{2\kappa}+i\epsilon)}\biggr]\;,\\
\chi^F(x(\tau),x(\tau'))&=&\frac{i}{4\pi\cos(\frac{i(\tau-\tau')}
     {2\kappa})}\delta'(\tau-\tau')\;.
 \eea
Performing the Fourier transform on the statistical functions and
then following the same procedure as that in the proceeding Section,
 we find that the Lamb shift for the static atom can be written as
 \beq
\Delta=\Delta_0+\Delta_{T_s}\label{total shift for static atom}
 \eeq
with
 \beq
\Delta_{T_s}=\frac{\mu^2}{8\pi^2}\int_{0}^{\infty}d\omega
         P\biggl(\frac{\omega}{\omega+\omega_{0}}
         -\frac{\omega}{\omega-\omega_{0}}\biggr)
         \frac{1}{e^{2\pi\kappa\omega}-1}\;.
 \eeq
Here $\Delta_{T_s}$ is the same as the correction to the Lamb shift
of an inertial atom  induced by the presence of a thermal bath at
the temperature
 \beq
T_s=\frac{1}{2\pi\kappa}=\frac{1}{2\pi\alpha\sqrt{g_{00}}}\;.\label{T
static}
 \eeq
 in a Minkowski spacetime. This differs from what was  obtained in the case of the
freely-falling atom $(T_f=1/2\pi\alpha)$. Remarkably, the two are
related by
 \beq
{T_s}^2=\biggl(\frac{1}{2\pi\alpha}\biggr)^2
     +\biggl(\frac{a}{2\pi}\biggr)^2=T^2_{f}+T^2_U\;\label{T^2_s}
 \eeq
with
 \beq
a=\frac{r}{\alpha^2}(1-\frac{r^2}{\alpha^2})^{-1/2}
 \eeq
being the inherent acceleration of the static atom. The first term
on the right hand side of Eq.~(\ref{T^2_s}) is the square of the
Gibbons-Hawking temperature of de Sitter spacetime. The second term,
which is acceleration-dependent, is a result of the Unruh effect.
So, the correction to the Lamb shift of the static atom  is a
combined effect of the thermal nature of de Sitter spacetime
characterized by the Gibbons-Hawking temperature  and  the Unruh
effect which is associated  with the inherent acceleration of the
atom. Thus, in terms of the Lamb shift, the temperature felt by the
static atom is the square root of the sum of the squared
Gibbons-Hawking temperature and the squared Unruh temperature
associated with its inherent acceleration. It is worthwhile to note
that the same relation as that of Eq.~(\ref{T^2_s}) is also obtained
in other different physical
contexts~\cite{Narnhofer96,Deser97,Zhiying08}.

\section{Summary}
Using the DDC formalism, we have calculated the contributions of
vacuum fluctuations and the radiation reaction to the energy level
shifts of both freely-falling and  static two-level atoms in
interaction with a conformally coupled massless real scalar field in
the de Sitter-invariant vacuum and obtained the Lamb shifts. The
Lamb shifts of both the freely-falling and the static atoms are in
structural similarity to that of an inertial atom immersed in a
thermal bath in the Minkowski spacetime.

For a freely-falling atom,  the Lamb shift gets a correction  as if
it was immersed in a thermal bath at the Gibbons-Hawking temperature
$T_f=1/2\pi\alpha$. This clearly reveals the intrinsic thermal
nature of de Sitter spacetime. The correction is however
insignificant in terms of actual experimental measurements, as the
cosmological constant is very tiny, corresponding to an effective
temperature as low as $\sim10^{-30}K$.

For a static atom, the Lamb shift is affected by a combination of
the Gibbons-Hawking effect of de Sitter spacetime and the Unruh
effect associated with the inherent acceleration of the atom. In
fact, the Lamb shift for a static atom in de Sitter spacetime is the
same as that of an inertial atom in a thermal bath in the Minkowski
spacetime at the temperature which is a square root of the sum of
the squared Gibbons-Hawking temperature and the squared Unruh
temperature associated with the atomic inherent acceleration.

\begin{acknowledgments}

We would like to thank the Kavli Institute for Theoretical Physics
China for hospitality where this work was finalized during a program
on AdS/CFT. This work was supported in part by the National Natural
Science Foundation of China under Grants No. 10775050, No. 11075083,
and No. 10935013; the Zhejiang Provincial Natural Science Foundation
of China under Grant No. Z6100077; the SRFDP under Grant No.
20070542002; the National Basic Research Program of China under
Grant No. 2010CB832803; the PCSIRT under Grant No. IRT0964; the
Project of Knowledge Innovation Program (PKIP) of Chinese Academy of
Sciences, Grant No. KJCX2.YW.W10;  and the Programme for the Key
Discipline in Hunan Province.

\end{acknowledgments}

\end{document}